# Optimized Fuzzy Logic Based Framework for Effort Estimation in Software Development

Vishal Sharma[1] and Harsh Kumar Verma[2]

[1] Department of Computer Science and Information Technology
DAV College, Jalandhar, Punjab, India

[2] Department of Computer Science and Engineering
Dr B R Ambedkar National Institute of Technology
Jalandhar, Punjab, India

**Abstract**
Software effort estimation at early stages of project development holds great significance for the industry to meet the competitive demands of today's world. Accuracy, reliability and precision in the estimates of effort are quite desirable. The inherent imprecision present in the inputs of the algorithmic models like Constructive Cost Model (COCOMO) yields imprecision in the output, resulting in erroneous effort estimation. Fuzzy logic based cost estimation models are inherently suitable to address the vagueness and imprecision in the inputs, to make reliable and accurate estimates of effort. In this paper, we present an optimized fuzzy logic based framework for software development effort prediction. The said framework tolerates imprecision, incorporates experts knowledge, explains prediction rationale through rules, offers transparency in the prediction system, and could adapt to changing environments with the availability of new data. The traditional cost estimation model COCOMO is extended in the proposed study by incorporating the concept of fuzziness into the measurements of size, mode of development for projects and the cost drivers contributing to the overall development effort.

## 1. Introduction

Software cost estimation is a vital aspect that guides and supports the planning of software projects. Controlling the expenses of software development effectively is of significant importance in today's competitive world [1, 2]. The need for reliable and accurate software development cost predictions in software engineering is a challenging perspective accounting for considerable financial and strategic planning [3]. Software cost estimation guides the prediction of the likely amount of effort, time, and staffing levels required to build a software system at an early stage during a project. However, estimates at the preliminary stages of the project are the most difficult to obtain because the primary source to estimate the costing comes from the requirement specification documents [4]. The accuracy of the estimates is quite low at the starting stages of the project because of the limited details available. Age old approaches for software projects effort prediction such as the use of mathematical formulae derived from historical data, or the use of expert's judgments, lack in terms of effectiveness and robustness in their results. These issues are even more critical when these effort prediction approaches are used during the early phases of the software development lifecycle (for instance, effort predictors along with their relationships to effort are characterized as being more imprecise and uncertain at requirements development phase, than those of later development phases, like design).

## 2. Effort Estimation Models

Software effort estimation stands as the oldest and most mature aspect of software metrics towards rigorous software measurement. Considerable research had been carried out in the past, to come up with a variety of effort prediction models. This section discusses the evolution of both algorithmic and non-algorithmic estimation techniques overtime.

2.1 Algorithmic Models

Algorithmic models predict effort relying upon the accurate estimate of either size of software in terms of lines of code (LOC), number of user screens, interfaces, complexity, etc. at a time when uncertainty is mostly present in the project [5]. Boehm was the first researcher to consider software engineering economically. He came up with a cost estimation model, COCOMO-81 in 1981, after investigating a large set of data from TRW in the 1970s assuming that the effort grows more than linearly on software size [6]. Putnam also developed an early model known as SLIM in 1978[7]. Both these models make use of data from past projects and are based on linear regression techniques, take number of lines of code (about which least is known very early in the project) as the major input to their models. A survey on these algorithmic models and other cost estimation approaches is presented by Boehm et. al.[8]. Algorithmic models such as COCOMO are unable to present suitable solutions that take into consideration technological advancements [3]. This is because, these models are often unable to capture the complex set of relationships (e.g. the effect of each variable in a model to the overall prediction made using the model) that are evident in many software development environments [9].They can be successful within a particular type of environment, but not flexible enough to adapt to a new environment. They cannot handle categorical data (specified by a range of values) and most importantly lack of reasoning capabilities. These limitations have paved way for the number of studies exploring non-algorithmic methods (e.g. Fuzzy Logic)[10].

## 2.2 Soft Computing Based Models

Newer computation techniques, to cost estimation that are non-algorithmic i.e. approaches that are soft computing based came up in the 1990s, and turned the attention of researchers towards them. This section discusses some of the non-algorithmic models for software development effort estimation. Soft computing encompasses methodologies centering in fuzzy logic (FL), artificial neural networks (ANN) and evolutionary computation (EC). These methodologies handle real life ambiguous situations by providing flexible information processing capabilities.

Soft computing techniques have been used by many researchers for software development effort prediction to handle the imprecision and uncertainty in data aptly, due to their inherent nature. The first realization of the fuzziness of several aspects of one of the best known [11], most successful and widely used model for cost estimation, COCOMO, was that of Fei and Liu [6]. They observed that an accurate estimate of delivered source instruction (KDSI) cannot be made before starting the project; therefore, it is unreasonable to assign a determinate number for it. Jack Ryder investigated the application of fuzzy modeling techniques to two of the most widely used models for effort prediction; COCOMO and the Function-Points models, respectively [5]. Fuzzy Logic was applied to the cost drivers of intermediate COCOMO model (the most widely used version) as it has relatively high estimation accuracy than the basic version which is quite comparable to the detailed version [12]. The study ignored the key project attribute "size" to estimate the software development effort. The resulting model lacked in one of the most desirable aspect of software estimation models i.e. adaptability. Musilek et al. applied fuzzy logic to represent the mode and size as input to COCOMO model [13]. The study was not adaptive as it lacked fuzzy rules which are definitely important to augment the system with expert's knowledge.

Ahmed et al. went a step further and fuzzified the two parts of COCOMO model i.e., nominal effort estimation and the adjustment factor. They proposed a fuzzy logic framework for effort prediction by integrating the fuzzified nominal effort and the fuzzified effort multipliers of the intermediate COCOMO model [10]. Knowing the likely size of a software product before it has been constructed is potentially beneficial in project management [14]. The results suggest that with refinement using data and knowledge, fuzzy predictive models can outperform their traditional regression-based counterparts.

Boetticher has described a neural network approach for characterizing programming effort based on internal product measures [15]. A study assessed the capabilities of a neuro-fuzzy system in comparison to other estimation techniques and models [3]. Neuro-fuzzy systems combine the valuable learning and modeling aspects of neural networks with the linguistic properties of fuzzy systems. An accuracy of within 25% of actual effort more than 75% of the time can be achieved for one large commercial data set for a neural network based model when used to estimate software development effort [16].

In summary, the previous research reveals that all of the soft computing-based software effort prediction models that exist, lack in some aspect or the other. There is still much uncertainty as to what prediction technique suits which type of prediction problem [17]. So, there is a compelling demand to develop a single soft computing based model which handles tolerance of imprecision in the input at the preliminary phases of software engineering, addresses the fuzzification of one of the key attribute i.e. size of the project, incorporates expert's knowledge in a well-defined manner, allows total transparency in the prediction system by prediction of results through rules or other means, adaptability towards continually changing development technologies and environments[18]. Properly addressing all these issues would position soft computing-based prediction techniques as models of choice for effort prediction, considering the promising features already present in them.

## 3. Optimized Fuzzy Logic Based Framework

This research developed an optimized fuzzy logic based framework to handle the imprecision and uncertainty present in the data at early stages of the project to predict the effort more accurately. The said framework is built upon an existing cost estimation model—COCOMO. The choice is justified in a way that, while many traditional models have been said to perform poorly when it comes to cost estimation, COCOMO-81 is said to be the best known [11], most plausible[13], and most cited [3] of all traditional models. The COCOMO model is a set of three models: basic, intermediate, and detailed [19]. This research used intermediate COCOMO model because it has estimation accuracy that is greater than the basic version, and at the same time comparable to the detailed version [12]. COCOMO model takes the following as input: (1) the estimated size of the software product in thousands of Delivered Source Instructions (KDSI) adjusted for code reuse; (2) the project development mode given as a constant value B (also called the scaling factor) ; and (3) 15 cost drivers [19, 20]. The development mode depends on one of the three categories of software development modes: organic, semi-detached, and embedded. It takes only three values, {1.05, 1.12, 1.20}, which reflect the difficulty of the development. The estimate is adjusted by factors called cost drivers that influence the effort to produce the software product. Cost drivers have up to six levels of rating: Very Low, Low, Nominal, High, Very High, and Extra High. Each rating has a corresponding real number (effort multiplier), based upon the factor and the degree to which the factor can influence productivity. The estimated effort in person-months (PM) for the intermediate COCOMO is given as:

$$\text{Effort} = A \times [KDSI]^B \times \prod_{i=1}^{15} EM_i \qquad (1)$$

The constant A in "(1)" is also known as productivity coefficient. The scale factors are based solely on the original set of project data or the different modes as given in Table 1.

Table 1: COCOMO Mode Coefficients and Scale Factor Values

| Mode | A | B |
| --- | --- | --- |
| Organic | 3.2 | 1.05 |
| Semidetached | 3.0 | 1.12 |
| Embedded | 2.8 | 1.2 |

The contribution of effort multipliers corresponding to the respective cost drivers is introduced in the effort estimation formula by multiplying them together. The numerical value of the ith cost driver is $EM_i$ and the product of all the multipliers is called the estimated adjustment factor (EAF).

The actual effort in person months (PM), $PM_{total}$ is the product of the nominal effort (i.e. effort without the cost drivers) and the EAF, as given in "(2)".

$$PM_{total} = PM_{nominal} \times EAF \qquad (2)$$

(where $PM_{nominal} = A \times [KDSI]^B$ and $EAF = \prod_{i=1}^{15} EM_i$)

The proposed framework addresses the limitations of existing soft computing based techniques for effort estimation by:

- Fuzzification of the two components of the COCOMO model (the nominal effort part and cost driver part) that capture imprecision in an organized manner.
- Incorporating expert's knowledge by providing a transparent and well defined approach by developing an appropriate rule base that can be modified.
- Integrating the two components of the COCOMO model viz. the nominal effort prediction component and the effort adjustment component.

The framework will thus allow fuzzy and expert knowledge incorporation into the system.

## 4. Research Methodology

Imprecision is present in all parameters of the COCOMO model. The exact size of the software project to be developed is difficult to estimate precisely at an early stage of the development process. COCOMO does not consider the software projects that do not exactly fall into one of the three identified modes. In addition, the cost drivers are categorical. Obviously, this limits the correctness and precision of estimates made. There is a need for a technology, which can overcome the associated imprecision residing within the final results of cost estimation. The technique endorsed here deals with fuzzy sets. In all the input parameters, fuzzy sets can be employed to handle the imprecision present.

In the course of research work the following steps were undertaken:

### 4.1 Choice of membership functions

Appropriate membership function representing the size of the project which is an input to the basic component (estimating nominal effort) of the underlying model i.e. COCOMO was identified. Gaussian membership functions proved superior to the triangular membership functions used in most of the previous researches for fuzzifying the sizes of projects, to address the vagueness in the project sizes. They are inherently adaptable, due to their nonlinearity and also allow a smoother transition in the intervals representing size as a linguistic variable as shown in "Fig. 1".

The same type of member ship functions are used for representing software development mode, to accommodate the projects falling between the identified modes as shown in "Fig. 2". In this way the framework can handle changing development environments, by accommodating projects that may belong partially to two categories of modes (80% semi- detached and 20% embedded). The resulting effort is also represented with Gaussian membership functions shown in "Fig. 3".

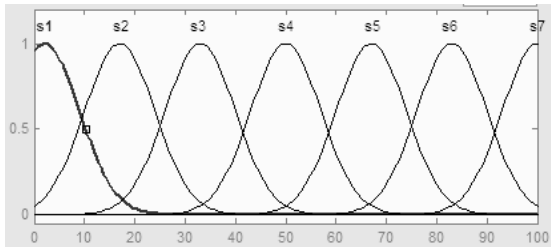

Fig. 1   Input variable "size" represented as Gaussian MF

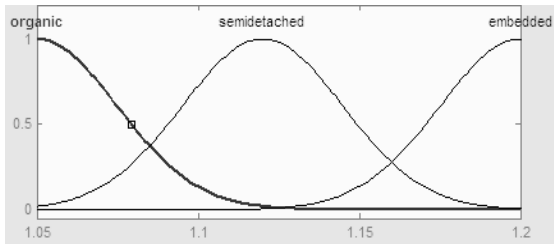

Fig. 2   Input variable "mode" represented as Gaussian MF

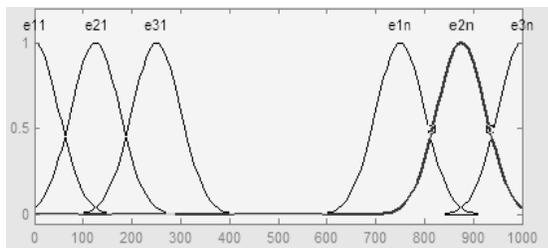

Fig. 3   Output variable "effort" represented as hypothetical Gaussian MF

4.2 Development of the fuzzy rules for nominal effort component

The basic component of the COCOMO model is used to develop the fuzzy rules to estimate nominal effort, independent of cost drivers thereby finding correspondence between mode, size and resulting effort by dividing input and output spaces into fuzzy regions [21, 22]. The parameters of the effort MFs were determined for the given mode, size pair. 3 MFs representing effort were obtained for a random size and 3 modes respectively. Rules formulated, based on the fuzzy sets of modes, sizes and efforts appear in the following form:

*IF mode is organic and size is s1 THEN effort is e11*
*IF mode is semi-detached and size is s1 THEN effort is e21*
*IF mode is embedded and size is s1 THEN effort is e31*
*IF mode is organic and size is s2 THEN effort is e12*
*IF mode is semi-detached and size is s2 THEN effort is e22*
*IF mode is embedded and size is s2 THEN effort is e32*
*.....*
*IF mode is mj and size is si THEN effort is eji*
$(1 \leq i \leq n, 1 \leq j \leq 3)$

where $m_j$ are the fuzzy values for the fuzzy variable mode, $s_i (1 \leq i \leq n)$ are the fuzzy values for the fuzzy variable.

4.3 Fuzzification of cost drivers

The cost drivers are fuzzified using triangular and trapezoidal fuzzy sets for each linguistic value such as very low, low, nominal, high etc. as applicable to each cost driver. Separate independent FIS is used for every cost driver. Rules are developed with cost driver in the antecedent part and corresponding effort multiplier in the consequent part. The defuzzified value for each of the effort multiplier is obtained from individual FISs after matching, inference aggregation and subsequent defuzzification. Total EAF is obtained after multiplying them together.

Sample fuzzification of main storage used (STOR) cost driver based on Tables 2 and 3 is illustrated in "Fig. 4 and 5".

Table 2: The STOR(Main Storage) Cost Driver Definition in terms of percentage

| *Nominal* | *High* | *Very High* | *Extra high* |
|---|---|---|---|
| <=50% | 70 | 85 | 95 |

Table 3: The STOR(Main Storage) Effort Multiplier Range Definitions

| *Nominal* | *High* | *Very High* | *Extra high* |
|---|---|---|---|
| 1.0 | 1.06 | 1.21 | 1.56 |

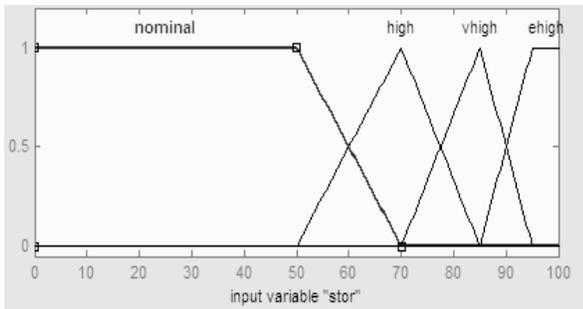

Fig. 4 Antecedent MFs for the FIS of cost driver main storage

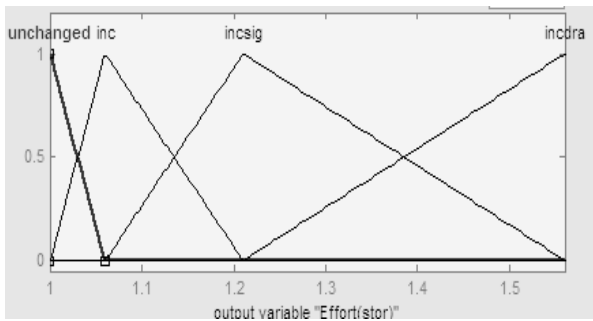

Fig. 5 Consequent MFs for the FIS of cost driver main storage

From "Figs. 4 and 5" rules of the following form are obtained:

*If stor is nom (nominal) Then Effort is unchanged*
*If stor is high Then Effort is inc (increased)*
*If stor is vhigh (very high) Then Effort is incsig (increased significantly)*
*If stor is ehigh (extra high) Then Effort is incdra (increased drastically)*

4.4 Integration of components

Total software effort is obtained by multiplication of crisp effort from the basic part and crisp EAF from the cost driver part (the product of effort multipliers corresponding to each of the 15 cost drivers).

In concluding the presentation of the framework, it is worth noting that rules are developed for the nominal effort part using COCOMO as the underlying model. The rules formulated for the cost drivers' are simply developed into FISs based on the tables in [21]. However, the membership functions definition and rules formulation are open to experts' knowledge, because our approach is transparent.

## 5. Experiments and Results

The approach has been validated by performing diverse experiments, on the proposed framework. COCOMO nominal equation has been used to generate artificial datasets randomly for developing the FIS for software effort prediction. The prediction capabilities of the FIS were tested using different numbers of fuzzy sets (3, 5, and 7) for input variable, size with triangular member-ship functions (TMFs) as well as Gaussian membership functions (GMFs). The performance of the FIS improved with the increased number of membership functions as shown in Figs. 6 and 7. The performance of the FIS is best when 7 MFs are used for size.

This is primarily due to a suitable rule base with respect to the fuzzy partitions of size, initially. This suggests that the number of fuzzy sets should be enough to cover the rules of the rule base appropriately and there is no under fitting or over fitting.

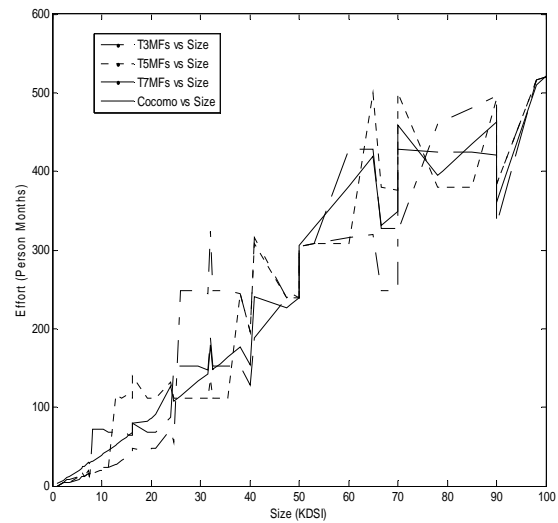

Fig. 6 Nominal Effort of FIS with 3,5,7 TMFs and COCOMO using COCOMO public dataset

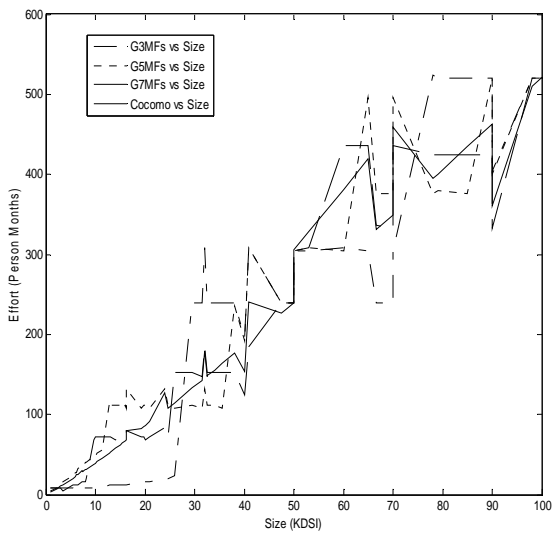

Fig. 7 Nominal Effort of FIS with 3,5,7 GMFs and COCOMO using COCOMO public dataset

The graph in Fig. 8 shows a comparison of nominal effort predicted by FIS using 7 triangular and 7 gaussian member-ship functions representing input variable, size against the nominal effort predicted by COCOMO. The experiments establish that gaussian membership functions perform better than triangular membership functions in terms of effort prediction.

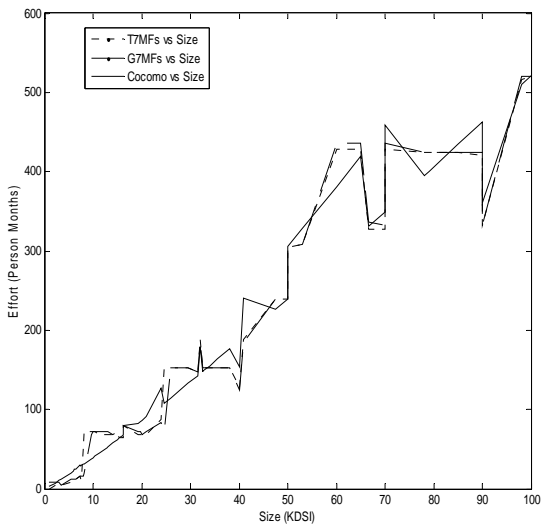

Fig. 8 Comparison of Nominal Effort of FIS with 7 TMFs, 7GMFs and COCOMO using COCOMO public dataset

A comparison is made for the mean magnitude of relative error (MMRE) in the estimate of nominal and total effort (adjusted with the effort multipliers) using 3, 5, and 7 membership functions for size against the prediction of COCOMO as shown in "Figs. 9 and 10".

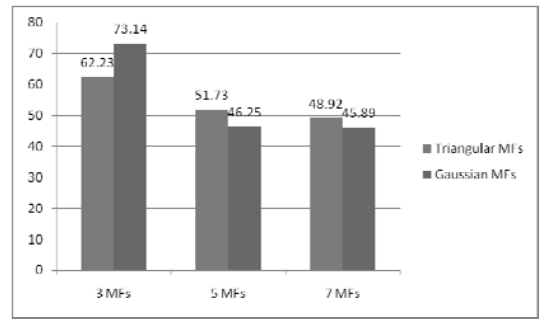

Fig. 9 MMRE in Nominal Effort predicted by FIS using 3, 5 and 7 TMFs and GMFs

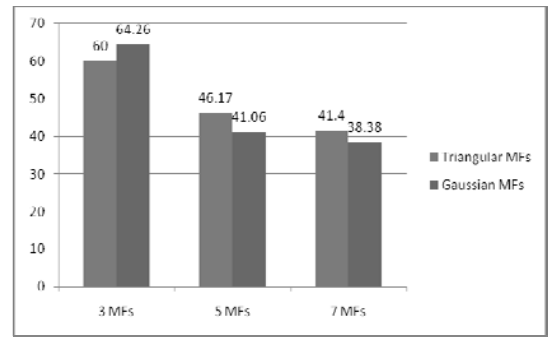

Fig. 10 MMRE in Total Effort predicted by FIS using 3, 5 and 7 TMFs and GMFs

The values of MMRE in nominal effort predicted is 39.6 using COCOMO model, 73.14, 46.25, and 45.89 using FIS with GMFs and 62.23, 51.73 and 48.92 using FIS with TMFs. The values of MMRE in total effort predicted is 38.83 using COCOMO model, 64.26, 41.06, and 38.38 using FIS with GMFs and 60, 46.17 and 41.4 using FIS with TMFs.

A quantitative measure of prediction quality is used to compare the prediction accuracy of our research approach to the actual values. The prediction quality of the FISs using 3, 5, and 7 member-ship functions is compared in terms of prediction accuracy both for nominal and total efforts. The quantity used in the experiments is prediction at level x – PRED(x). Suppose there is a set of *n* projects, let y be the number of them, whose mean magnitude of relative error is less than or equal to x, then:

$$PRED(x) = y/n \qquad (3)$$

An acceptable level for mean magnitude of relative error is something less than or equal to 0.25. The prediction accuracy of FISs in terms of predicted nominal effort and total using 3, 5, and 7 TMFs and GMFs for size is shown in Table 4. The results suggest that GMFs outperform TMFs in terms of effort prediction within 25% of the actual effort, when used to represent size and mode.

Table 4: Readings of PRED (25) of FIS

| Number of MFs | *Prediction Accuracy PRED(25) using Triangular MFs for Effort* | | *Prediction Accuracy PRED(25) using Gaussian MFs for Effort* | |
|---|---|---|---|---|
| | *Nominal* | *Total* | *Nominal* | *Total* |
| 3 | 16.92 | 15.38 | 15.38 | 18.46 |
| 5 | 20 | 33.84 | 32.3 | 41.54 |
| 7 | 33.84 | 41.54 | 35.38 | 43.07 |

The validation experiments are carried out on the FIS using COCOMO public dataset. The validation was performed using a subset of the real life projects whose size fall in the range 1-100 KDSI. This is justified because the original FIS was developed over the same range of size using artificial datasets generated using nominal effort component of COCOMO.

The comparison of nominal effort prediction by FIS and intermediate COCOMO model is made on actual real project data and is shown graphically in "Fig. 11".This validation is of extreme importance because it measures the prediction quality of the proposed framework against the actual real life data of software development projects.

Another validation experiment compares the comprehensive effort predicted by FIS and intermediate COCOMO model with the incorporation of effort multipliers as given in the actual real project data. It is shown graphically "Fig. 12".

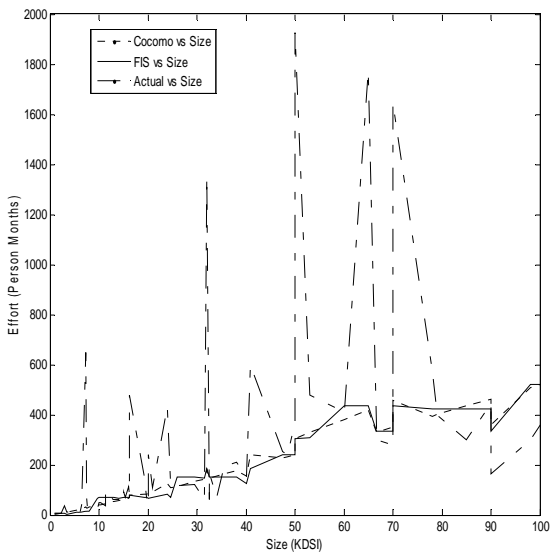

Fig. 11 Nominal Effort predicted by FIS, COCOMO and Actual effort using COCOMO public dataset

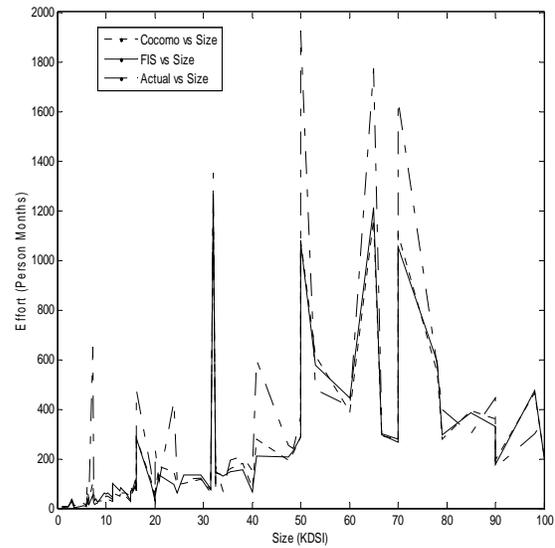

Fig. 12 Total Effort predicted by FIS, COCOMO and Actual effort using COCOMO public dataset

A comparison is made between the quality of prediction results by plotting the percentage error in nominal effort predicted by the FIS and nominal effort estimated by the COCOMO model against the sizes of the projects as given in the COCOMO database. This is represented graphically by "Fig. 13". The results indicate that the percentage errors in the nominal effort predicted by the FIS are less than 50% for most of the projects except a few.

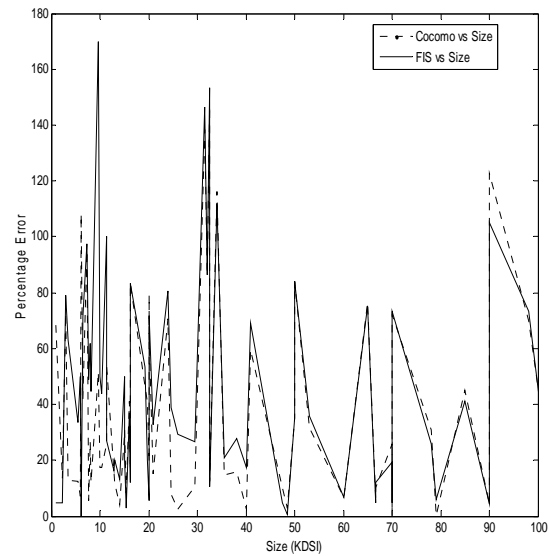

Fig. 13 Percentage Error in nominal effort predicted by FIS and COCOMO using COCOMO public dataset

Another comparison between the comprehensive efforts taking into account the cost drivers as well in terms of percentage errors in both the proposed FIS and COCOMO model is shown in "Fig. 14". The errors fall in the range of 0-50% for most of the projects and even better the prediction of age old COCOMO model except for one project as indicated in the graph. This may be attributed to the fact that the FIS has been developed using COCOMO as the underlying model.

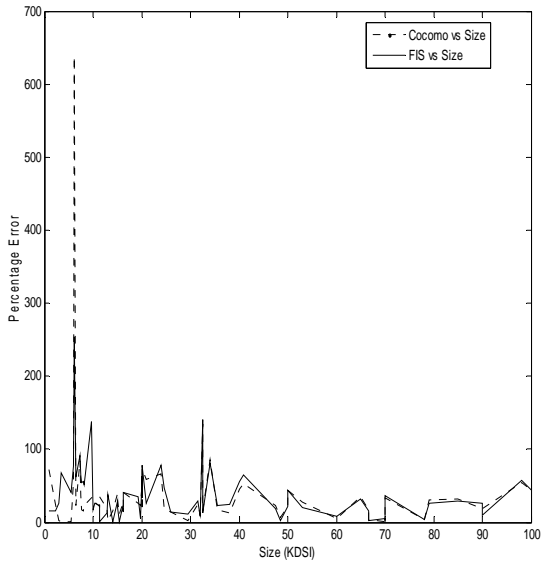

Fig. 14 Percentage Error in total effort predicted by FIS and COCOMO using COCOMO public dataset

## 6. Conclusions and Future Scope

The research presents a transparent, optimized fuzzy logic based framework for software development effort prediction. The Gaussian MFs used in the framework have shown good results by handling the imprecision in inputs quite well and also their ability to adapt further make them a valid choice to represent fuzzy sets. The framework is adaptable to the changing environments and handles the inherent imprecision and uncertainties present in the inputs quite well. The framework is augmented by the contribution of the experts in terms of modifiable fuzzy sets and rule base in accordance with the environments. The performance of the framework is demonstrated in terms of empirical validation carried on live project data of the COCOMO public database. However, a little more insight into the training strategies and availability of more real life data suggest a room for improvement in the prediction results.

This research indicates directions for further research. Some of the identified future motivations of research are as follows:

1) The proposed framework can be analyzed in terms of feasibility and acceptance in the industry.

2) The framework can be deployed on COCOMO II environment with experts providing required information for developing fuzzy sets and an appropriate rule base.

3) With a little more knowledge in fuzzy logic, customized MFs can be developed to represent inputs more closely to tolerate imprecision and uncertainty in inputs so that the same is not propagated to the outputs.

4) Novice training strategies can be incorporated to normalize the error measure..